\documentclass[
reprint,
superscriptaddress,
nofootinbib,
nobibnotes,
amsmath,amssymb,
aps,
pra,
showkeys,
]{revtex4-1}

\usepackage{graphicx}
\usepackage{bm}
\usepackage{amsfonts}
\usepackage[normalem]{ulem}
\usepackage{wasysym}
\usepackage{siunitx}
\usepackage{url}
\usepackage{color}

\newcommand*{\diff}{\,\mathrm{d}}

\begin{document}

\title{A first search for a stochastic gravitational-wave background\\
	from ultralight bosons}

\author{Leo Tsukada}
\email{tsukada@resceu.s.u-tokyo.ac.jp}
\affiliation{
	Research Center for the Early Universe (RESCEU), Graduate School of Science, \\The University of Tokyo, Tokyo 113-0033, Japan}
\affiliation{Department of Physics, Graduate School of Science, The University of Tokyo, Tokyo 113-0033, Japan}
\author{Thomas Callister}
\affiliation{LIGO Laboratory, California Institute of Technology, Pasadena, CA 91125, USA}

\author{Andrew Matas}
\affiliation{School of Physics and Astronomy, University of Minnesota, Minneapolis, Minnesota 55455, USA}
\affiliation{Max Planck Institute for Gravitational Physics (Albert Einstein Institute), D-14476 Potsdam-Golm, Germany}

\author{Patrick Meyers}
\affiliation{School of Physics and Astronomy, University of Minnesota, Minneapolis, Minnesota 55455, USA}
\affiliation{OzGrav, University of Melbourne, Parkville, Victoria 3010, Australia}

\date{\today}

\begin{abstract}
	Ultralight bosons with masses in the range $ \SI{e-13}{eV}\leq m_b\leq\SI{e-12}{eV}$ can induce a superradiant instability around spinning black holes (BHs) with masses of order 10-100 $M_\odot$. This instability leads to the formation of a rotating ``bosonic cloud'' around the BH, which can emit gravitational waves (GWs) in the frequency band probed by ground-based detectors. 
	The superposition of GWs from all such systems can generate a stochastic gravitational-wave background (SGWB). In this work, we develop a Bayesian data analysis framework to study the SGWB from bosonic clouds using data from Advanced LIGO and Advanced Virgo, building on previous work by Brito \textit{et.al.} \cite{brito}. We further improve this model by adding a BH population of binary merger remnants. To assess the performance of our pipeline, we quantify the range of boson masses that can be constrained by Advanced LIGO and Advanced Virgo measurements at design sensitivity. Furthermore, we explore our capability to distinguish an ultralight boson SGWB from a stochastic signal due to distant compact binary coalescences (CBC). Finally, we present results of a search for the SGWB from bosonic clouds using data from Advanced LIGO's first observing run. We find no evidence of such a signal. Due to degeneracies between the boson mass and unknown astrophysical quantities such as the distribution of isolated BH spins, our analysis cannot robustly exclude the presence of a bosonic field at any mass. Nevertheless, we show that under optimistic assumptions about the BH formation rate and spin distribution, boson masses in the range $ \SI{2.0e-13}{eV}\leq m_\mathrm{b}\leq\SI{3.8e-13}{eV} $ are excluded at 95\% credibility, although with less optimistic spin distributions, no masses can be excluded. The framework established here can be used to learn about the nature of fundamental bosonic fields with future gravitational wave observations.
\end{abstract}

\keywords{gravitational waves, stochastic backgrounds, superradiant instability, Bayesian inference}
\maketitle

\section{\label{sec:1}Introduction}
	The first detection of gravitational waves (GWs) from binary black hole (BBH) \cite{GW150914} and neutron star (BNS) \cite{GW170817} coalescences by the Advanced Laser Interometer Gravitational-wave Observatory (LIGO) \cite{TheLIGOScientific:2014jea} and Advanced Virgo \cite{TheVirgo:2014hva} represented a historical breakthrough, creating an alternative window through which to view the Universe. Furthermore, subsequent observations of BBH coalescence events have firmly established GW astronomy \cite{GW151226, GW170104, GW170814, GW170608, O2_catalog}, shedding new light on fundamental physics \cite{TheLIGOScientific:2016src} and the nature of the stellar mass black hole (BH) population \cite{TheLIGOScientific:2016htt, O2_pop}. In the near future, Advanced LIGO and Advanced Virgo will be joined by additional detectors, like KAGRA \cite{kagra} and LIGO-India \cite{ligo-india}. A major target for the advanced detector network is the stochastic gravitational-wave background (SGWB), a superposition of many sources too faint to resolve individually \cite{GW150914_implication,GW170817_implication}.
	
	Ultralight bosons around spinning BHs have been proposed as a possible source of the SGWB \cite{brito, brito_short}. Superradiant instabilities induced by the bosonic fields co-rotating around BHs result in angular momentum extraction analogous to the so-called Penrose process \cite{penrose}. This instability not only spins down the BH, but also triggers an exponential growth of the bosonic field. Subsequently the resulting bosonic cloud exhibits a time-dependent quadrupole moment, leading to GW radiation. We could obeserve these GWs with current ground-based GW detectors in two regimes -- the ``resolvable'' regime in which nearby sources within $\mathcal{O}(10\,{\rm Mpc})$ can be directly detected \cite{brito, semicoherent, CW_search, lilly} and the ``unresolvable'' regime where a superposition of all other sources in the Universe will contribute to a SGWB \cite{brito, brito_short}.
	
	Ground-based detectors are sensitive to bosons with a given mass scale that can be determined through dimensional-analysis \cite{brito, brito_short}. We expect a bosonic field with the mass $ m_b $ to couple strongly to BHs whose Schwarzschild radius $r_H$ is comparable to the boson's Compton wavelength $\lambda_C=\hbar/m_bc$, where $ \hbar $ is the reduced Planck constant and $ c $ is the speed of light. For BHs of around $ 10M_{\odot} $, this implies
	
	\begin{align}
		\label{eq:Mmu1}
		m_bc^2&\sim \frac{\hbar c^3}{2GM}\sim\SI{e-13}{\electronvolt}\times\left(\frac{M}{\SI{10}{\textit{M}_{\odot}}}\right)^{-1},
	\end{align}
	where $ G $ is the gravitational constant and $M$ is the mass of a BH. 
	In other words, only BHs within a relatively narrow window will significantly couple to bosonic field with a certain mass scale. This can be physically interpreted as follows. For a given boson mass, if BH mass is too large or small, the timescale of superradiant instability increases and its effect is exponentially suppressed \cite{brito, brito_review, Dolan}. As we will demonstrate, Advanced LIGO could detect superradiant instabilities surrounding $\sim 10 M_\odot$ BHs, and therefore can probe boson masses near $m_b \approx \SI{e-13}{\electronvolt}$.
	
	Light bosonic fields, such as QCD axions or axion-like particles have attracted attention, as a promising candidate of dark matter \cite{string_axiverse, dark}. It is challenging to use traditional particle physics experiments to probe the existence of light particles that do not couple strongly to ordinary matter, but we can use GW observations to study particle physics beyond the Standard Model \cite{Arvanitaki1, Arvanitaki2}.
	
	Since angular momentum extraction is followed by characteristic GW emission, the detection of this class of GWs could also provide an explanation for the possible abundance of low spin BHs \cite{superradiance}, which is consistent with the current BH spin measurement by Advanced LIGO \cite{O1_catalog, GW170814, GW170608, O2_pop}. Additionally, the BH spin distribution is considered to be a crucial indicator of different BH formation channels \cite{Belczynski2016, isolate_binary, dynamical1, dynamical2}. The discovery of ultralight bosons regulating BH spins would have significant implications for understanding BH formation scenarios. 
	
	Previous work has studied the SGWB from bosonic clouds \cite{xilong, brito_short}. Ref.~\cite{xilong} examined the detectability of a stochastic background from BHs spinning down, assuming the current ground-based detector network operates at design sensitivity. They assumed two different signal spectra, referred to as the Gaussian model and the quasi-normal-mode model, and using a Fisher analysis they also assessed the capability of extracting the component of their signal models from the total background including the CBC background. Ref.~\cite{brito_short} computed the background spectrum expected from the superradiant instability assuming only an isolated BH population model. They also compared the predicted spectrum with the power-law integrated curves \cite{PI_curve} of Advanced LIGO to study its detectability. In this work, we will develop a Bayesian analysis framework to search for the SGWB from the superradiant instability and apply it to data from Advanced LIGO's first observing run. We construct our signal model based on \cite{brito, brito_short}, additionally incorporating the expected contribution from remnant BHs produced by compact binary mergers to obtain a more accurate prediction.
	
	This paper is structured as follows. In Section \ref{sec:2} we review the model of superradiant instabilities as previously computed in \cite{brito, brito_short, brito_review}. In Section \ref{sec:3}, we present the predicted SGWB signal, including the contribution from population of binary merger remnants. In Section \ref{sec:4}, we present a Bayesian framework with which to search for this background. In Section \ref{sec:5}, we discuss the sensitive range in the boson mass and the model selection capability between this background and the projected SGWB that arises due to unresolved CBCs. Furthermore, in Section~\ref{sec:O1_zerolag} we show the range of excluded boson masses using data from Advanced LIGO's first observing run. Finally, Section \ref{sec:6} summarizes the results and future work.
	
	In what follows, all quantities are described in units $ G=c=1 $.
\section{\label{sec:2}Superradiant instability}
	Before describing the data analysis methods, first we briefly review the theoretical basis of the bosonic cloud model as described in \cite{brito,brito_short}. For a review of supperradiance, see \cite{superradiance}. Hereafter, we will restrict ourselves to scalar (spin-0) fields. Although vector (spin-1) \cite{sr_vector_2, sr_vector_3, sr_vector_4} and tensor (spin-2) \cite{sr_tensor} fields can additionally induce superradiant instabilities, the scalar case has been investigated most extensively \cite{Arvanitaki1, Arvanitaki2, Dolan, kodama_yoshino_2014, superradiance}.
	
	Incident radiation can by amplified by extracting energy and angular momentum from a BH. As a result, the mass and angular momentum of the BH transitions from the initial states $ M_i, J_i $ to the final states $ M_f, J_f $, respectively. Denoting the frequency of the radiation by $\omega$ and letting $l,m$ be the quantum numbers for angular momentum, the condition for superradiance to occur is given by \cite{superradiance}
	\begin{align}
		\label{eq:superradiant}
		0<\omega<m\Omega_H,
	\end{align}
	where $ \Omega_H $ is the horizon angular velocity of the BH, given by
	\begin{align}
		\Omega_H=\frac{\chi}{2r_+}.
	\end{align}
	Here, $ \chi $ is the dimensionless spin parameter and $ r_+ $ is the outer event horizon of a Kerr BH. 
	
	Using black hole perturbation theory, in the limit that the scalar field's Compton wavelength is much larger than the radius of the BH ($ M_i\mu \ll 1 $), where 
	\begin{align}
		\mu\equiv \frac{m_b}{\hbar},
	\end{align}
	an approximate expression can be found for the eigenfrequencies of the scalar field \cite{Dolan, brito}: 
	\begin{align}
		\begin{split}
			\label{eq:omega_lmn}
			\omega_{lmn}&\equiv\omega_R+i\omega_I\\
			&\simeq \mu +i~2\gamma \mu r_+(m\Omega_H-\mu)(M_i\mu)^{4l+4},
		\end{split}
	\end{align}
	where $n,l,m$ are the quantum numbers for energy and angular momentum for a wave function in a spherical potential, $ \gamma $ is a positive numerical factor depending on $ n, l, m $, and $\omega_{R/I}$ are real quantities. The imaginary part of the eigenfrequency, $\omega_I$, will be positive, indicating that the mode is (linearly) unstable, whenever Eq. \eqref{eq:superradiant} is satisfied\footnote{Note that we have used the relation $\omega_R=\mu$, which is valid to leading order in $M_i\mu$.}. This instability leads to the formation of a scalar ``cloud'' around the BH. We define the instability timescale $\tau_\mathrm{inst}\equiv\omega_I^{-1}$ as the characteristic timescale over which the cloud forms. We will focus only on the $n=0$, $l=m=1$ mode hereafter since this mode has the shortest timescale and dominates the gravitational-wave radiation. In the large Compton wavelength limit, the instability timescale is given by \cite{brito_short, brito_review, brito} 
	\begin{align}
		\label{eq:t_inst}
		\tau_\mathrm{inst}\sim0.07\chi^{-1}\left(\frac{M_i}{\SI{10}{\textit{M}_{\odot}}}\right)\left(\frac{0.1}{M _i\mu}\right)^{9} \mathrm{yr}.
	\end{align}
	\begin{figure*}[t]
		\centering
		\includegraphics[width=0.8\textwidth]{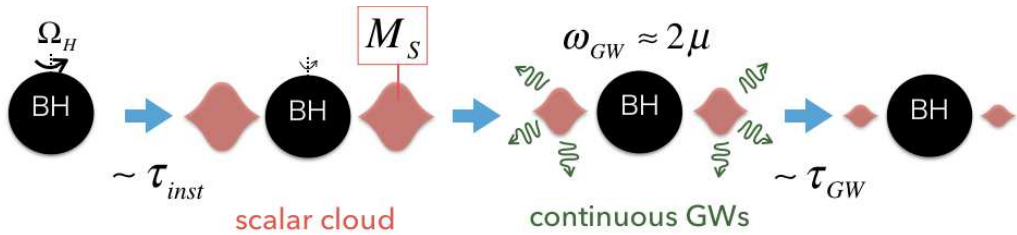}
		\caption{\label{fig:sr_inst_cartoon} Cartoon of a superradiant instability. $ M_S $ is the total mass of amplified scalar field. The angular frequency $ \omega_{GW}$ of emitted GWs depends on the mass of the scalar field $ m_b $ (or equivalently the inverse of its Compton wavelength $\mu$). The timescale of the instability and GW emission are denoted by $ \tau_{\mathrm{inst}} $ and $ \tau_{\mathrm{GW}} $, respectively. If Eq.~\eqref{eq:superradiant} is satisfied, quantum vacuum fluctuations in the scalar field will repeatedly scatter off the BH, growing exponentially with a typical timescale $ \tau_\mathrm{inst}\sim\SI{0.07}{\mathrm{yr}}$ (see Eq.~\eqref{eq:t_inst}). The result of this runaway process is a scalar cloud, whose quadrupole moment induces gravitational radiation at the constant frequency $ \omega_{\mathrm{GW}}=2\mu$. For BHs with a mass of order 10 $M_\odot$, GWs are emitted continuously over a typical timescale $ \tau_\mathrm{GW}\sim\SI{6e4}{\mathrm{yr}} $ (see Eq.~\eqref{eq:t_gw}) until the energy of the scalar cloud has been completely dissipated.}
	\end{figure*}
	
	As illustrated in Figure~\ref{fig:sr_inst_cartoon}, after the scalar cloud has been formed, the scalar field profile will source a stress-energy tensor with a time-dependent quadrupole moment. The energy stored in the rotating scalar field configuration will emit gravitational-waves, with a frequency of $\omega_{\rm GW}=2\omega_R$ and a luminosity given by \cite{brito_short, brito_review, brito} 
	\begin{align}
		\label{eq:dE_dt}
		\frac{\diff  E_{\mathrm{GW}}}{\diff  t}\approx\frac{484+9\pi^2}{23040}\left(\frac{M_S}{M_f}\right)^2(M_f\mu)^{14},
	\end{align}
	where $ M_S $ is the total mass (energy) of the amplified scalar field configuration. Eventually, the scalar cloud will lose energy due to gravitational radiation. This process will happen over the course of the emission timescale $\tau_{\rm GW}$, given by \cite{brito_short, brito_review, brito} 
	\begin{align}
		\label{eq:t_gw}
		\tau_{\mathrm{GW}}\sim 6\times10^4 \chi^{-1}\left(\frac{M_f}{\SI{10}{\textit{M}_{\odot}}}\right)\left(\frac{0.1}{M_f\mu}\right)^{15} \mathrm{yr}.
	\end{align}
	
	Crucially, there is a significant difference between the instability and emission timescales, $ \tau_{\mathrm{GW}}/\tau_{\mathrm{inst}}\sim10^6 $. Thus, we make the approximation that the scalar cloud forms more or less instantaneously and then starts to emit GWs. Neglecting the emission of GWs during the growth of the scalar cloud, conservation of energy and angular momentum implies that \cite{brito_short}
	\begin{align}
		\label{eq:Jf_1}
		J_f=J_i - \frac{m}{\omega_R}(M_i-M_f).
	\end{align}
	The amplification of the cloud stops when the condition $\omega_R=m\Omega_H$ is satisfied. Combining this with Eq.~\eqref{eq:Jf_1}, we obtain the final mass of the BH 
	\begin{align}
		M_f=\frac{m^3-\sqrt{m^6-16m^2\omega_R^2(mM_i-\omega_RJ_i)^2}}{8\omega_R^2(mM_i-\omega_RJ_i)}.
	\end{align}
	For the lowest mode $m=1$, in the long Compton wavelength limit $M_i\mu\ll 1$, the maximum mass of the scalar cloud reads
	\begin{align}
		M_S^{\mathrm{max}}=M_i-M_f \approx 0.1M_i\chi\left(\frac{M_i\mu}{0.1}\right).
	\end{align}
	Thus, for typical parameters $M_i \mu\approx 0.1$, about 10\% of the energy of the initial BH is stored in the scalar cloud configuration. 
	
	Finally, we can evaluate the total GW energy emitted between the superradiance saturation and the present time as follows
	\begin{align}
		\label{eq:E_gw}
		E_\mathrm{GW}&=\int_{0}^{\Delta t}\diff  t\frac{\diff  E_\mathrm{GW}}{\diff t}=\frac{M_S^\mathrm{max}\Delta t}{\Delta t+\tau_\mathrm{GW}},
	\end{align}
	where $ \tau_\mathrm{GW} $ is the GW emission timescale. Here, the signal duration $ \Delta t $ is reasonably defined as $ \Delta t \equiv\min(\tau_\mathrm{GW}, t_0) $, in which $ t_0 $ is the age of the Universe $ \approx\SI{13.8}{Gyr} $. The SGWB will be calculated by summing the energy emitted by each source over the population of rotating BHs.
\section{\label{sec:3}Modeling the SGWB}
	Now we turn to modelling the SGWB produced by scalar clouds which form around BHs. We consider isolated BHs, following the model of \cite{brito}. We also derive the contribution due to remnant BHs formed from compact binary coalescences, and verify that it is smaller than the contribution of isolated BHs for most scalar masses of interest.
	
	\subsection{Framework to compute the SGWB}
	The SGWB is described by its energy density spectrum \cite{allen_joe, joe_neil}, defined as
	\begin{align}
		\label{omega_gw}
		\Omega_\mathrm{GW}(f)\equiv\frac{1}{\rho_c}\frac{\mathrm{d}\rho_{\mathrm{GW}}}{\mathrm{d}\ln(f)},
	\end{align}
	where $ \rho_{\mathrm{GW}} $ is the energy density of GWs existing in the Universe and $ \rho_c $ is the critical energy density required to have a spatially flat Universe. The astrophysical SGWB has been studied for a wide array of sources \cite{CBC_form_1, CBC_form_2, CBC_form_3, CBC_form_4, string_form_1, string_form_2} and is generally given by
	\begin{align}
		\label{eq:omgwf_gen}
		\Omega_\mathrm{GW}(f)=\frac{f}{\rho_c}\int\diff z\frac{\diff t}{\diff z}\int\diff \boldsymbol{\theta}p(\boldsymbol{\theta})R(z; \boldsymbol{\theta})\frac{\diff E_s}{\diff f_s}(\boldsymbol{\theta};f(1+z)),
	\end{align}
	where $ \diff E_s / \diff f_s $ is the source-frame energy spectrum of an individual astrophysical event, $ R(z; \boldsymbol{\theta}) $ is the number of sources per unit comoving volume per unit \textit{source frame} time, and $p(\boldsymbol{\theta}) $ is the multivariate probability distribution of the source parameters $ \boldsymbol{\theta} $.
	Note that the source frame frequency $f_s$ is related to the observed frequency $f$ by a factor of the redshift: $f_s=(1+z)f $. The function ${\rm d}t/{\rm d} z$ is determined by standard cosmology
	\begin{align}
		\frac{\diff t}{\diff z} = \frac{1}{(1+z)H_0\sqrt{\Omega_M(1+z)^3+\Omega_{\Lambda}}},
	\end{align}
	where $ \Omega_M$ and $  \Omega_\Lambda$ are the dimensionless matter density and the dimensionless cosmological constant density, respectively. We use the cosmological parameters inferred from Planck \cite{planck}, i.e. $ H_0 = \SI{68}{\kilo\metre ~\second^{-1} Mpc^{-1}}$ and $ \Omega_M=1-\Omega_{\Lambda}=0.308 $ .
	
	When constructing a model for the SGWB from a superradiant instability, we take the source parameters $ \boldsymbol{\theta} $ to be BH masses and their dimensionless \textit{initial} spin parameter $\chi$. This $ \chi $ represents the BH spin before spinning down due to the instability. We note that the GWs emitted from an individual system actually have a slight frequency drift caused by the change in the cloud's binding energy during the GW emission \cite{lilly}. However, its overall drift is well within the width of each frequency bin that we use for our analysis. This leads to the following simple spectrum for a single source
	\begin{align}
		\frac{\diff E_\mathrm{s}}{\diff f_s}\approx E_\mathrm{GW}\delta(f(1+z)-f_0),
	\end{align}
	where $ f_0=\omega_R/\pi\approx\mu/\pi $ (see Eq.~\eqref{eq:omega_lmn}) and $ E_\mathrm{GW} $ is given by Eq.~\eqref{eq:E_gw}.
	
	In order to compute the background due to unresolved sources, we only integrate over the range of parameters $\theta$ that produces a signal-to-noise ratio (SNR) of less than 8 in a typical search for the astrophysical system in question \cite{CW_search, lilly}. Nevertheless, the specifics of this cutoff do not change the overall shape of the predicted spectra significantly. 
	
	\subsection{BH population models}
	To search for and constrain the stochastic background from a superradiant instability, we will need to assume a specific source number density and the mass and spin distributions of BHs. Below, we will consider two possible BH populations: (i) isolated BHs formed by core-collapse supernovae (CCSNe) and (ii) BBH merger remnants.
	
	\subsubsection{\label{sec:iso_model} Isolated black holes}
	Here we assume a population of isolated BHs born from CCSNe. Eq.~\eqref{eq:omgwf_gen} can be reorganized as
	\begin{align}
		\label{eq:iso_omega_gw}
		\Omega_\mathrm{GW}^\mathrm{iso}(f)=\frac{f}{\rho_c}\int\diff z\frac{\diff t}{\diff z}\int\diff M\diff\chi p(\chi)\frac{\diff \dot{n}}{\diff M}\frac{\diff E_s}{\diff f_s},
	\end{align}
	where $ {\diff\dot{n}}/{\diff M} $ is the source-frame BH formation rate per comoving volume per BH mass $M$. For the probability density of $ \chi $, we assume a uniform distribution
	\begin{align}
		p(\chi) =
		\begin{cases}
			0&(\chi<\chi_\mathrm{ll}, \chi_\mathrm{ul}<\chi)\\
			\frac{1}{\chi_\mathrm{ul}-\chi_\mathrm{ll}}&(\chi_\mathrm{ll}\leq\chi\leq\chi_\mathrm{ul}),
		\end{cases}
	\end{align}
	where $ \chi_\mathrm{ll}, \chi_\mathrm{ul} $ are the lower and upper limit of the distribution. Inspired by \cite{brito_short}, we adopt two different parameterizations of $p(\chi)$, either (a) varying the lower limit $\chi_\mathrm{ll}$ and fixing $\chi_\mathrm{ul}=1$, or (b) varying the upper limit $\chi_\mathrm{ul}$ and fixing $\chi_\mathrm{ll}=0$. Note that the first case is more optimistic than the second case, as it ensures a population of high-spin BHs that readily yield superradiant instabilities. Although these models for $p(\chi)$ admittedly quite simple, the true spin distribution of isolated BHs is extremely uncertain. As we will show, different $ \chi_{\mathrm{ll,ul}} $ values crucially affect the background spectrum. 
	
	The BH formation rate $ \diff \dot{n}/\diff M $ reads
	\begin{align}
		\frac{\diff \dot{n}}{\diff M}&\equiv p(M)R(z;M)\\
		\label{eq:iso_rate}
		&=\psi(z_f)\frac{\xi(\mathcal{M}_*)}{\mathcal{M}_*}\frac{\diff \mathcal{M}_*}{\diff M}.
	\end{align}
	$ \mathcal{M}_* $ is the mass of the BH's progenitor star, whose population properties follow the cosmic star formation rate (SFR) $ \psi(z) $ and initial mass function (IMF) $ \xi( \mathcal{M}_*) $. We adopt the SFR model proposed in \cite{SFR},
	\begin{gather}
		\begin{split}
			\label{eq:SFR}
			\psi(z)=\nu\frac{a\exp(b(z-z_m))}{a-b+b\exp(a(z-z_m))},\\
			a=2.37, b=1.80, \nu=0.178, z_m=2.00.
		\end{split}
	\end{gather}
	$ z_f $ is the redshift at the progenitor's time of birth, that is, $ t_f= t-\tau(\mathcal{M}_*)$, where $ \tau(\mathcal{M}_*) $ is the lifetime of a progenitor based on \cite{lifetime}. Note that $ \xi(\mathcal{M}) $ is the IMF defined in terms of stellar mass fraction. This implies that $ \xi(\mathcal{M}_*)\diff \mathcal{M}_*$ yields the ratio of the total mass of stars whose mass lies between $ \mathcal{M}_*\sim\mathcal{M}_*+\diff \mathcal{M}_*$ to the whole stellar mass. Since the Salpeter IMF is chosen as $ \xi(\mathcal{M}_*) $ in the mass range $ \mathcal{M}_*\in[0.1\sim100]~M_\odot $, its expression is\footnote{One should not confuse it with the other definition, that is, \textit{the number fraction} of stars whose mass lies between $ \mathcal{M}_*\sim\mathcal{M}_*+\diff \mathcal{M}_*$. Let this be $ \phi(\mathcal{M}_*) $. The Salpeter function in this definition follows $ \phi(\mathcal{M}_*)\propto \mathcal{M}_*^{-2.35}$.}
	\begin{align}
		\label{eq:xi}
		\xi(\mathcal{M}_*)=\frac{\mathcal{M}_*^{-1.35}}{\int_{0.1M_\odot}^{100M_\odot}\mathcal{M}_*^{-1.35}\diff \mathcal{M}_*}.
	\end{align}
	The BH mass is related to the mass and metalliticity of the progenitor star via $ M=g(\mathcal{M}_*, Z) $. In this work we use a numerical fit for $g(\mathcal{M}_*, Z)$ given by \cite{g_M}. It follows that
	\begin{align}
		\frac{\diff  \mathcal{M}_*}{\diff M}= \left(\frac{\diff g(\mathcal{M}_*, Z)}{\diff \mathcal{M}_*}\right)^{-1}.
	\end{align}
	Note that $ g(\mathcal{M}, Z) $ implicitly depends on redshift via the stellar metallicity $Z$, whose evolution over the cosmic history is evaluated in \cite{metallicity}. The lower BH mass cutoff of $ g(\mathcal{M}, Z) $ is set as $ \SI{3}{\textit{M}_\odot} $.
	
	\subsubsection{\label{sec:BBH_model} Binary black hole merger remnants}
	We also consider remnant BHs formed by BBH mergers, whose background spectrum is evaluated such that
	\begin{align}
		\begin{split}
			\label{eq:bbh_omega_gw}
			\Omega_\mathrm{GW}^\mathrm{rem}(f)=\frac{f}{\rho_c}&\int\diff z\frac{\diff t}{\diff z}\\
			\times&\int\diff \boldsymbol{m}\diff \chi p\,(\boldsymbol{m})R_m(z;\boldsymbol{m})p(\chi)\frac{\diff E_s}{\diff f_s}.
		\end{split}
	\end{align}
	$ \boldsymbol{m} $ represents a set of component masses of BBHs, i.e. the primary mass $ m_1 $ and the secondary mass $ m_2 $ ($ m_1>m_2 $ by definition) and $ \diff\boldsymbol{m} \equiv \diff m_1\diff m_2$. $ R_m(z; \boldsymbol{m}) $ is the BBH merger rate density for a given $ \boldsymbol{m}$ and $ z $. Unlike the spin distribution of isolated BHs, here $ p(\chi) $ can be motivated by both the spin measurement of final remnant BHs by Advanced LIGO and Virgo \cite{O1_catalog, GW170104, GW170608, GW170814} and several numerical simulations for mergers of similar-mass BHs \cite{final_spin_1, final_spin_2, final_spin_3}. Both suggest that the spin magnitude of final remnant BHs is around $ 0.7 $. Therefore, we assume all the remnant BHs initially have $ \chi=0.7 $, namely
	\begin{align}
		\label{eq:rem_pchi}
		p(\chi) = \delta(\chi-0.7).
	\end{align}
	
	Following \cite{GW170817_implication, O1_iso, O1_catalog}, we adopt the BH mass distribution
	\begin{align}
		p(m_1, m_2) \propto \frac{m_1^{-2.35}}{m_1-5M_{\odot}},
	\end{align}
	with the constraints that $ 5M_{\odot}<m_1, m_2<95M_{\odot} $ and $ m_1 + m_2<100M_{\odot} $. We approximate the mass of remnant BHs as
	\begin{align}
		M\approx m_1 + m_2 - 5.7\times10^{-2}\frac{m_1m_2}{m_1+m_2},
	\end{align}
	where the binding energy of the inner most stable orbit in a binary system is subtracted from the total mass \cite{Maggiore}.
	
	We evaluate the merger rate as described in \cite{Dominik_2,GW150914_implication, GW170817_implication, O1_iso},
	\begin{align}
		\label{eq:merger_rate}
		R_m(z;\boldsymbol{m}) = \int_{t_\mathrm{min}}^{t_\mathrm{max}}R_f(z_f;\boldsymbol{m})p(t_d)\diff t_d,
	\end{align}
	where $ t_d $ is the time delay between a binary formation and its merger and $ p(t_d) $ is the time delay distribution. $ z_f $ is the redshift at the binary formation time $ t_f = t(z) - t_d $, in which $ t(z) $ is the cosmic time at merger. Following \cite{GW150914_implication, GW170817_implication}, we assume that: (i) the distribution $ p(t_d) $ is taken as $ p(t_d) \propto 1/t_d$ in the range $ t_\mathrm{min}<t_d<t_\mathrm{max} $, where $ t_\mathrm{min} $ is $ 50~\mathrm{Myr} $ \cite{GW150914_implication, Dominik_2} and $ t_\mathrm{max} $ is the Hubble time \cite{nakar_tmin, SGRB_tmin, Dominik_1, Dominik_2}, (ii) the formation rate of BBHs, both of which are lighter than $ 30M_{\odot} $, evolves proportionally to the SFR $ \psi(z) $, (iii) the SFR of \cite{SFR} is consistent with the one used for the isolated BHs model (see Eq.~\eqref{eq:SFR}), (iv)  massive binaries in which at least one of the two component masses is above $ 30M_{\odot} $ cannot be formed in a high metallicity environment where $ Z > Z_{\odot}/2 $ ($ Z_{\odot} $ is the solar metallicity). For the assumption (iv), we instead multiply the SFR by a weighting factor $ e(z) $ to account for the fraction of star formation in an environment where $ Z \leq Z_{\odot}/2 $ \cite{Madau_2014}. In other words,
	\begin{align}
		\label{eq:R_f}
		R_f(z_f; \boldsymbol{m})\propto
		\begin{cases}
			\psi(z_f)&\mathrm{if}\;m_1, m_2<30M_{\odot}\\
			e(z_f)\psi(z_f)&\mathrm{otherwise}.
		\end{cases}
	\end{align}
	After marginalizing Eq.~\eqref{eq:merger_rate} over the component mass distribution $ p(\boldsymbol{m}) $, the calibration at $ z=0 $ is performed with the published estimate of local merger rate with the power-law mass distribution\footnote{At the time of writing the paper, Advanced LIGO and Virgo collaboration published a new paper on the inferred BBH population property using their first and second observing runs and the local merger rate estimate has been updated. However, we stick to the local merger rate, previously estimated in \cite{GW170104}, to demonstrate the pipeline performance. The normalization with the new merger rate estimate will be considered in future work.}, $ 103~\mathrm{Gpc}^{-3}\mathrm{yr}^{-1} $ \cite{GW170104} such that
	\begin{align}
		\label{eq:rateNorm}
		\int p(\boldsymbol{m}) R_m(z=0;\boldsymbol{m})\diff\boldsymbol{m}  = 103~\mathrm{Gpc}^{-3}\mathrm{yr}^{-1}.
	\end{align}
	
	Although we have assumed this rate is known exactly, we note that the true rate is uncertain -- the current 90\% credible bounds on the BBH merger rate span [9.7, 101]$~\mathrm{Gpc}^{-3}\mathrm{yr}^{-1}$. This uncertainty will correspondingly introduce a systematic uncertainty in our prediction of the remnant spectrum $\Omega^\mathrm{rem}_\textsc{gw}(f)$, whose effect should be considered in future work.
	
	\subsection{Total background model}
\begin{figure}
		\centering
		\includegraphics[width=\linewidth]{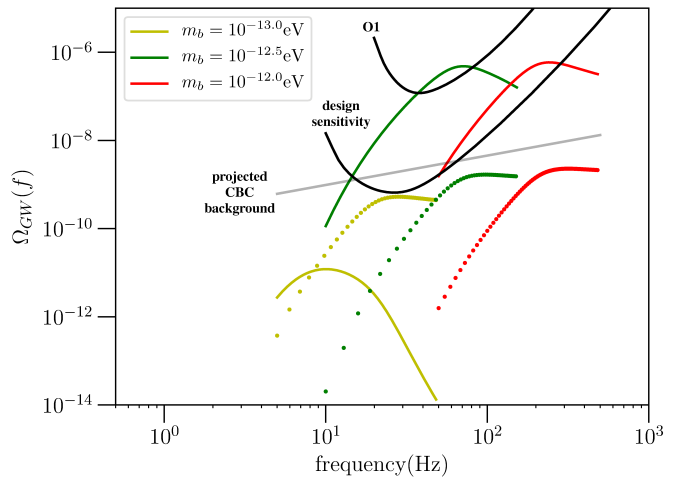}
		\caption{\label{fig:omega_gw_sum} Energy density spectra in the LIGO band overlapped with the power-law integrated curves \cite{PI_curve} of LIGO O1 \cite{O1_release} and design sensitivity \cite{PI_design}. Solid curves are spectra based on the isolated BH model with uniform distribution of $ \chi \in [0, 1.0] $, whereas dotted curves represent spectra with the BBH merger remnant model. The gray line indicates the projected background of compact binary coalescence (CBC) modeled as a simple power-law spectrum with a power law index of $ 2/3 $ \cite{GW170817_implication}. The solid yellow curve is much lower than the other curves, because of the predicted lack of isolated BHs with large enough mass to couple to scalar fields with $m_b=10^{-13}{\ \rm eV}$.}
	\end{figure}

	The actual background spectrum that one would observe is the sum of the contributions from these two BH populations, that is,
	\begin{align}
		\label{eq:SI_sgwb}
		\Omega_\mathrm{GW}(f) = \Omega_\mathrm{GW}^\mathrm{iso}(f) + \Omega_\mathrm{GW}^\mathrm{rem}(f),
	\end{align} 
	where the superscripts represent each of the isolated BH and BBH merger remnant population given by Eq.~\eqref{eq:iso_omega_gw} and Eq.~\eqref{eq:bbh_omega_gw}, respectively. For the comparison between these components, Fig.~\ref{fig:omega_gw_sum} shows background spectra computed from each BH population model, plotted with power-law integrated curves \cite{PI_curve} of Advanced LIGO at different phases and the CBC background approximated as a power-law spectrum \cite{GW170817_implication}. 
	
	The solid curves are the energy density spectra contributed from the isolated BH population for different scalar masses under an assumption of a uniform distribution $ p(\chi) $ over \SIrange{0}{1}{}.
	The spin distribution skewed towards high $ \chi $ would simply scale the overall spectrum.
	See Fig.~2 of \cite{brito_short} for its dependence on different spin distributions.
	The dotted curves, meanwhile, show the contribution from BBH merger remnants.
	In general, the abundance of isolated BHs exceeds the merger rate by four orders of magnitude.
	The isolated BH channel therefore dominates the SGWB for scalar masses $m_b\geq10^{-12.5}\mathrm{eV}$.
	However, if the $m_b\sim10^{-13}\mathrm{eV}$, the SGWB from isolated BHs is significantly suppressed due to the lack of BHs heavier than $50\,M_\odot$.
	In other words, although scalar fields with $ m_\mathrm{b}\sim\SI{e-13}{eV} $ is sensitively coupled to these heavy BHs (see Fig.~1 of \cite{brito}), such massive BHs are only rarely produced by CCSNe \cite{g_M}.
	In contrast, heavy BHs can readily be produced by the merger remnant channel, and so remnant BHs dominate the SGWB when $m_b\sim10^{-13}\mathrm{eV}$.
	
	It is worth noting that the SGWB may dominate the projected CBC background, which is approximated as
	\begin{align}
		\label{eq:omega_CBC}
		\Omega_\mathrm{GW}^\mathrm{CBC}(f) = 1.8\times10^{-9}\left(\frac{f}{25\mathrm{Hz}}\right)^{2/3},
	\end{align}
	as inferred from \cite{GW170817_implication}. We will discuss the capability to distinguish between these two signal models based on the Bayesian framework in Section~\ref{sec:ms}.
\section{\label{sec:4}Search Method}
	This section provides the overview of a statistical methodology that can be used to claim a detection or to place constraints related to the model of the superradiant instability. Here, we assume the GW background to be; (a) isotropic, (b) unpolarized, (c) stationary and (d) Gaussian.
	
	\subsection{Setup}
	The SGWB is analyzed by taking the cross-correlation between outputs from a pair of detectors. In this work we will focus on the case of two detectors for simplicity, but the formalism can be extended to handle a larger network of detectors \cite{allen_joe,joe_neil}. Following the notation of \cite{tom_nonGR_method}, we define a cross-correlation estimator for each frequency bin \cite{sgwb_hanford, joe_neil} as,
	\begin{align}
		\label{eq:estimator_def}
		\hat{C}(f) \equiv \frac{f^3}{T}\frac{20\pi^2}{3H_0^2}\tilde{s}_1^*(f)\tilde{s}_2(f).
	\end{align}
	Here, $ \tilde{s}_i(f) $ is the Fourier transform of time series output of the $ i $-th detector and $ T $ is the total observation time. This is normalized such that
	\begin{align}
		\label{eq:estimator_mean}
		\left\langle \hat{C}(f)\right\rangle =\gamma(f) \Omega_\mathrm{GW}(f),
	\end{align}
	where $ \gamma(f) $ is the overlap reduction function to encode the geometry and separation between a pair of detectors \cite{ORF}. 
	In the low signal-to-noise limit, the variance is approximately given by
	\begin{align}
		\label{eq:sigma_method}
		\sigma^2(f)\approx\frac{1}{2T\Delta f}\left(\frac{10\pi^2f^3}{3H_0^2}\right)^2P_1(f)P_2(f),
	\end{align}
	where $ \Delta f $ is frequency resolution and $ P_i(f) $ is the power spectral density (PSD) of the $ i $-th detector. 
	
	Then, the unbiased broadband estimator and its variance can be constructed from $ \hat{C}(f)$ and $\sigma^2(f)$ as follows
	\begin{align}
		\hat{Y}\equiv\frac{\sum_{f}\hat{C}(f)w(f)/\sigma^2(f)}{\sum_fw(f)/\sigma^2(f)}
	\end{align} 
	and
	\begin{align}
		\sigma_Y^2=\frac{\sum_{f}w^2(f)/\sigma^2(f)}{\left(\sum_fw(f)/\sigma^2(f)\right)^2},
	\end{align}
	where $ w(f) $ is a linear filter formed by a given background model $ \Omega_\mathcal{A}(f) $, so that
	\begin{align}
		w(f)=\gamma(f)\Omega_\mathcal{A}(f).
	\end{align}
	Once the broadband estimator is obtained, the SNR can be computed in a straightforward manner.
	\begin{align}
		\label{eq:SNR_def}
		\mathrm{SNR}\equiv\frac{\hat Y}{\sigma_Y}.
	\end{align}

	\subsection{\label{sec:bayes}Bayesian inference}
	We now develop a Bayesian formalism to perform parameter estimation and model selection for the superradiant instability model, following \cite{PE_stochastic}. In practice, the GW signal can be described by a set of unknown parameters $ \boldsymbol{\theta}=\{\theta_1, \theta_2 ..., \theta_N\} $ and one needs to evaluate their posterior probability given new data. Then, Bayes' theorem states that
	\begin{align}
		\label{eq:bayes_post}
		p(\boldsymbol{\theta}_\mathcal{A}|\{\hat{C}\}, \mathcal{A})=\frac{L(\{\hat{C}\}|\boldsymbol{\theta}_\mathcal{A}, \mathcal{A})\pi(\boldsymbol{\theta}_\mathcal{A}|\mathcal{A})}{Z(\{\hat{C}\}|\mathcal{A})},
	\end{align}
	where $ \{\hat{C}\} $ is the cross-correlation estimator over frequency band computed from observed data and $ \boldsymbol{\theta}_\mathcal{A} $ is a set of parameters characteristic of GW signals in the signal hypothesis $ \mathcal{A} $. $p(\boldsymbol{\theta}_\mathcal{A}|\{\hat{C}\}, \mathcal{A})$ is the posterior probability on the multi-dimensional space,  $L(\{\hat{C}\}|\boldsymbol{\theta}_\mathcal{A}, \mathcal{A})$ is the likelihood,  $\pi(\boldsymbol{\theta}_\mathcal{A}|\mathcal{A})$ is the prior probability and $Z(\{\hat{C}\}|\mathcal{A})$ is the evidence. To construct a posterior in an efficient manner, we apply the \texttt{PyMultiNest} package \cite{pymultinest} to our search pipeline. \texttt{PyMultiNest} is a python interface to the nested sampling package \texttt{MultiNest} \cite{multinest1, multinest2, multinest3}, which produces a set of samples drawn from an estimated posterior.
	
	\subsubsection{Likelihood}
	For one realization of $ \hat{C}(f) $, the joint likelihood over frequency bins is given by \cite{PE_stochastic}
	\begin{align}
		\label{eq:L_f}
		L(\{\hat{C}\}|\boldsymbol{\theta}_\mathcal{A}, \mathcal{A})=\prod_{f}L(\hat{C}(f)|\boldsymbol{\theta}_\mathcal{A}, \mathcal{A}).
	\end{align}
	$ L(\hat{C}(f)|\boldsymbol{\theta}_\mathcal{A}, \mathcal{A}) $ is the likelihood within a single frequency bin. We assume a Gaussian likelihood, such that
	\begin{align}
		\label{eq:L_nb}
		\begin{split}
			\ln&\left[L(\hat{C}(f)|\mathcal{\theta}_\mathcal{A}, \mathcal{A})\right] \equiv\\
			&-\frac{\left[\hat{C}(f)-\gamma(f)\Omega_\mathcal{A}(f;\boldsymbol{\theta}_\mathcal{A})\right]^2}{2\sigma^2(f)}-\frac{1}{2}\ln\left(2\pi\sigma^2(f)\right).
		\end{split}
	\end{align}
	Here, $ \Omega_\mathcal{A}(f;\boldsymbol{\theta}_\mathcal{A}) $ is a model energy-density spectrum for a given set of parameters $ \boldsymbol{\theta}_\mathcal{A} $.
	
	\subsubsection{Posterior}
	We choose a uniform prior on every parameter (e.g. scalar mass $m_b$ and spin limits $\chi_{\mathrm{ul}/\mathrm{ll}}$):
	\begin{align}
		\pi(\boldsymbol{\theta}_\mathcal{A}|\mathcal{A})\propto1.
	\end{align}
	Our posteriors are therefore proportional to the likelihood:
	\begin{align}
		p(\boldsymbol{\theta}_\mathcal{A}|\{\hat{C}\}, \mathcal{A}) \propto L(\{\hat{C}\}|\boldsymbol{\theta}_\mathcal{A}, \mathcal{A}).
	\end{align}
	We will often consider the marginalized posterior for a particular parameter, $\theta_1$, which is defined as
	\begin{align}
		p(\theta_1|\{\hat{C}\}, \mathcal{A})=\int p(\boldsymbol{\theta}_\mathcal{A}|\{\hat{C}\}, \mathcal{A})\mathrm{d}\theta_2...\mathrm{d}\theta_N.
	\end{align}
	The marginalized posterior will allow us to define credible intervals for the parameter $\theta_1$ in a standard way.
	
	\subsubsection{Model selection}
	In order to perform model selection between different models of the SGWB, we need to compute the Bayesian evidence for each hypothesis. Let $ \{\hat{C}\} $ be the cross-correlation estimator obtained from the data and suppose that one assesses which model, $ \mathcal{A}$ or $\mathcal{B}$, is better supported by the data. It is straightforward to compute the odds ratio $\mathcal{O}^{\mathcal{A}}_{\mathcal{B}}$, which is defined as
	\begin{align}
		\mathcal{O}_\mathcal{B}^\mathcal{A}\equiv\frac{p(\mathcal{A}|\{\hat{C}\})}{p(\mathcal{B}|\{\hat{C}\})}=\frac{Z(\{\hat{C}\}|\mathcal{A})}{Z(\{\hat{C}\}|\mathcal{B})}\frac{\pi(\mathcal{A})}{\pi(\mathcal{B})}.
		\label{eq:ms_BF}
	\end{align}
	where $ Z(\{\hat{C}\}|\mathcal{A})$ and $Z(\{\hat{C}\}|\mathcal{B}) $ are the evidences for each model. The evidence $ Z $ is calculated through
	\begin{align}
		Z(\{\hat{C}\}|\mathcal{A})=\int L(\{\hat{C}\}|\boldsymbol{\theta}_\mathcal{A}, \mathcal{A})\pi(\boldsymbol{\theta}_\mathcal{A}|\mathcal{A})\mathrm{d}^D\boldsymbol{\theta}_\mathcal{A}.
	\end{align}
	This expression can also be interpreted as the fully-marginalized likelihood. In the case where no signal is present (the null hypothesis $H_0$), the evidence is obtained by fixing $\Omega_{\mathcal{A}}(f;\theta_\mathcal{A})$ to zero. Also $Z(\{\hat{C}\}|\mathcal{A})/Z(\{\hat{C}\}|\mathcal{B})$ in Eq.~\eqref{eq:ms_BF} is the so-called Bayes factor, while $\pi(\mathcal{A})/\pi(\mathcal{B})$ is \textit{a priori} probability ratio for the two models, which we will set as unity. Since in our framework the odds ratio Eq.~\eqref{eq:ms_BF} is effectively equivalent to the Bayes factor, hereafter we will evaluate statistical significance in terms of the Bayes factor and follow the convention that a log Bayes factor of $ \approx8 $ indicates a favor for a model over the other with great confidence \cite{Jeffreys}.
\section{\label{sec:5}Results}
	In this section, we first describe the parameterization of our signal model. Section~\ref{sec:injection} shows a scheme for injecting simulated signals into Gaussian noise and recovering those signals. We then discuss the implications of performed analysis tests, including the construction of a ``sensitivity window'' of scalar masses to which Advanced LIGO might be sensitive. In Section~\ref{sec:ms}, we examine prospects for successfully discriminating between a CBC background and a SGWB due to supperradiant instability.
	
	\subsection{Parameters for the background model}
	In this work, the signal model is parameterized by the scalar mass $ m_b $ and either $ \chi_{\mathrm{ll}} $ or $ \chi_{\mathrm{ul}} $ (refer to Sections~\ref{sec:iso_model} for the details).
	We note that this model contains several sources of systematic uncertainty, such as the BH formation and merger rates as well as the underlying SFR.
	Nevertheless, we will assume a particular SFR (Eq. \eqref{eq:SFR}) and binary formation rate (Eq. \eqref{eq:rateNorm}). 
	Our results are also conditional on our specific parameterization of the $p(\chi)$. A different choice for the BH spin distribution will generically yield different constraints on $m_b$. Also, posteriors will be constructed over the two-dimensional parameter space of $m_b$ and either $\chi_\mathrm{ll}$ or $\chi_\mathrm{ul}$.
	
	\subsection{\label{sec:injection} Injection scheme and signal recovery}
	Here we describe the overview of the injection scheme adopted in this work. We assume only the LIGO detector pair, that is, the Hanford and Livingston sites. Injections are performed in the frequency domain.
	Given a PSD of the LIGO detectors, one constructs the variance of a cross-correlation estimator given by Eq.~\eqref{eq:sigma_method}.
	A cross-correlation spectrum consistent with Gaussian noise is then constructed from this variance and the SGWB predicted from superradiant instability is added to the Gaussian noise. Therefore, the simulated cross-correlation spectrum is defined as 
	\begin{align}
		\label{eq:C_inj}
		\hat{C}_\mathrm{sim}(f)\equiv\gamma_{\mathrm{HL}}(f)\Omega_{\mathrm{inj}}(f;\boldsymbol{\theta}_\mathrm{inj})+\sigma(f)\hat{n},
	\end{align}
	where $ \gamma_{\mathrm{HL}}(f)\ $ is the overlap reduction function for the LIGO baseline \cite{ORF}, $  \Omega_\mathrm{inj}(f;\boldsymbol{\theta}_\mathrm{inj}) $ is the injected background constructed from the parameters $ \boldsymbol{\theta}_\mathrm{inj} $, and $ \hat{n} $ is a random variable drawn from a Gaussian distribution with zero mean and unit variance. From this simulated cross-correlation spectrum, our pipeline infers the parameters $ \boldsymbol{\theta}_\mathrm{inj} $ by computing the likelihood and constructing a Bayesian posterior.
	We adopt a uniform prior for $m_b$ across the range \SIrange{e-13}{e-12}{eV} and a uniform prior for $ \chi_\mathrm{ul} $ or $ \chi_\mathrm{ll}$ between 0 and 1.
	
	\begin{figure}
		\includegraphics[width=\linewidth]{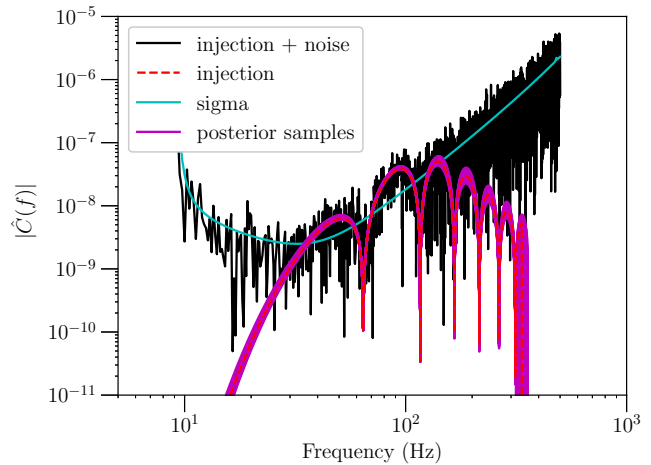}
		\caption{\label{fig:inj_estimator} An example of a cross-correlation spectrum $\hat{C}_\mathrm{sim}(f)$ as a function of frequency (the black line), obtained from a simulated observation of a SGWB from superradiant instabilities. The cyan curve is the standard deviation of $\hat{C}_\mathrm{sim}(f)$ calculated from the design sensitivity after two years of observation. The red dashed line gives the injected signal. The injection has SNR $ \sim $ 32 and corresponds to the superradiant instability model with $ m_\mathrm{b} =\SI{7e-13}{eV}$ and $ \chi_\mathrm{ll}=0.5$. The magenta region represents the model evaluated with 100 sets of the parameters randomly drawn from the posterior.}
	\end{figure}
	
	\begin{figure}
		\includegraphics[width=\columnwidth]{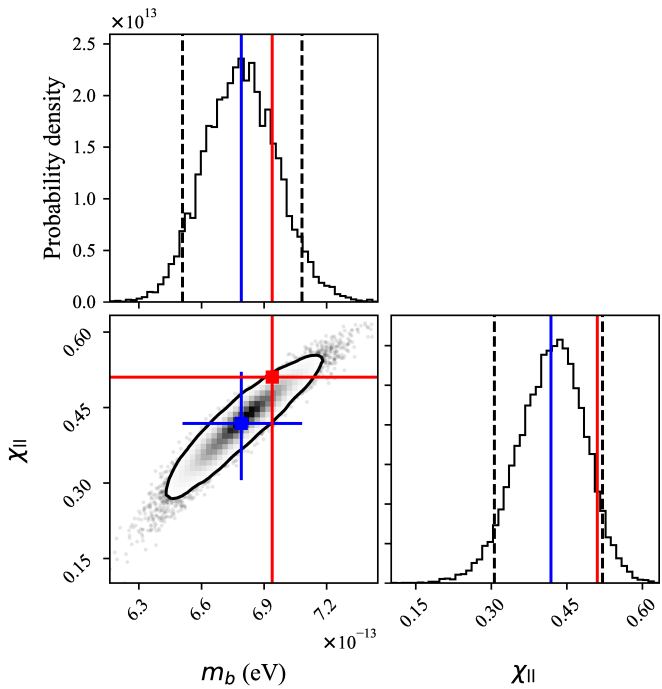} 
		\caption{\label{fig:inj_post_ul} Posterior result on scalar mass $m_b$ and spin lower limit $\chi_\mathrm{ll}$, given the simulated observation shown in Fig. \ref{fig:inj_estimator}. Within each one-dimensional posterior, the mean is depicted by a vertical blue solid line and marginalized $ 90\% $ credible interval is shown by two dashed lines. On the two-dimensional posterior, the blue cross is drawn for the mean and the credible interval and the black contour represents $ 90\% $ credibility. Also, the true injected parameters are indicated by the red cross. The errors in parameter recovery are consistent with Gaussian noise; see the consistency tests reported in Appendix \ref{app_pp}.}
	\end{figure}
	
	Fig.~\ref{fig:inj_estimator} shows an example cross-correlation spectrum $\hat{C}_\mathrm{sim}(f)$ as a function of frequency using two years of observation with Advanced LIGO's design sensitivity.
	The red dashed line gives the injected signal. The injection has SNR $ \sim $ 32 and corresponds to the superradiant instability model with $ m_\mathrm{b} =\SI{7e-13}{eV}$ and $ \chi_\mathrm{ll}=0.5$. The magenta region represents the model evaluated with 100 sets of the parameters randomly drawn from the posterior.
	Fig. \ref{fig:inj_post_ul} shows the resulting posterior on $m_b$ and $\chi_\mathrm{ll}$ derived from this simulated observation.
	These figures both demonstrate that the pipeline can recover a loud injection with great confidence. A statistical discussion for a consistency test of the parameter estimation is described in Appendix~\ref{app_pp}.
	
	\subsection{\label{sec:sw} Sensitivity window}
\begin{figure}
		\centering
		\includegraphics[width=\linewidth]{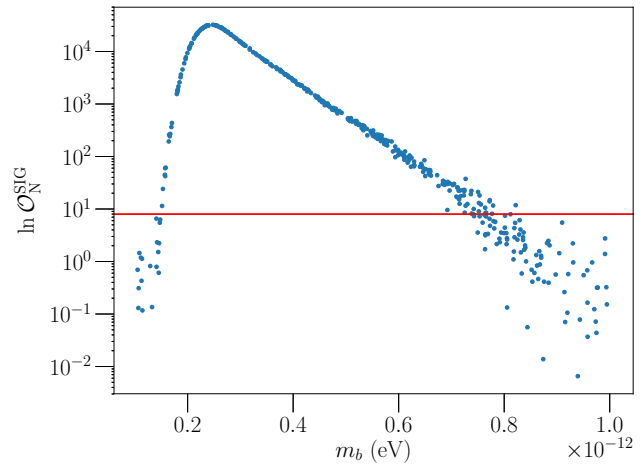}
		\caption{\label{fig:BF_plot} Bayes factors as a function of injected $ m_b $ values. The red horizontal line indicates a log Bayes factor of 8. In this test, we conservatively adopt the $ \chi_\mathrm{ul} $ parameterization with $ \chi_\mathrm{ul} $ of 0.8.}
	\end{figure}

	To study the scalar mass space we can probe through this model, which we refer to as the ``sensitivity window'', we make a number of injections and compute their Bayes factors between the signal and noise hypotheses.
	The signal injection is performed in the same way as Eq.~\eqref{eq:C_inj}. For this test, we conservatively adopt the $ \chi_\mathrm{ul} $ parameterization and the injected $ \chi_\mathrm{ul} $ value is fixed as $ 0.8 $. The $ m_b $ value for each of the 500 injections is uniformly drawn from the range \SIrange{.e-13}{.e-12}{eV}. In Fig.~\ref{fig:BF_plot}, we recognize the injections above a log Bayes factor of 8 as detectable. Therefore, the $ m_b $ range in which a log Bayes factor is larger than 8 (presented as the red line in Fig.~\ref{fig:BF_plot}) can be interpreted as the sensitivity window, which is around \SIrange{1.8e-13}{7.5e-13}{eV} in Fig.~\ref{fig:BF_plot}. Note that the actual sensitivity window depends on spin parameter $\chi_\mathrm{ul} $. Since larger $ \chi_\mathrm{ul} $ simply scales the overall spectrum by a constant factor, we can have a wider detectable range of scalar masses for a larger $ \chi_{\mathrm{ul}} $.

	\subsection{\label{sec:ms} Distinguishing superradiant instability and CBC backgrounds}
	
	As seen in Fig.~\ref{fig:omega_gw_sum}, the SGWB signal from the superradiant instability model dominates over the projected CBC background for some choices of scalar mass and BH spins. Thus, the next question to address is ``Given a detected signal, can we distinguish these two models from one another?''
	Here, we consider the case where both the superradiant instability and the projected CBC background are present, so that
	\begin{align}
		\label{eq:ms_sgwb_model}
		\Omega_\mathrm{inj}(f; \boldsymbol{\theta}) = \Omega_\mathrm{inj}^\mathrm{SI}(f; \boldsymbol{\theta}) + \Omega_\mathrm{inj}^\mathrm{CBC}(f),
	\end{align}
	where $\Omega_\mathrm{inj}^\mathrm{SI}(f; \boldsymbol{\theta})$ is the background due to superradiant instabilities under the $ \chi_{\mathrm{ul}} $ parametrization given by Eq.~\eqref{eq:SI_sgwb}, and $\Omega_\mathrm{inj}^\mathrm{CBC}(f) $ is the fixed CBC background approximated as the power law spectrum shown in Eq.~\eqref{eq:omega_CBC}.
	Note that, after three years of observation with Advanced LIGO's design sensitivity, $ \Omega_\mathrm{inj}^\mathrm{CBC}(f) $ alone is detectable with a log Bayes factor of 8.8 between a CBC-only and a noise model, which corresponds to an SNR of 5.5.
	
	Given a simulated measurement of the combined background shown in Eq.~\eqref{eq:ms_sgwb_model}, we recover our measurements with different two models: a CBC-only model and a joint superradiant instabiltity and CBC model (SI$ + $CBC). The CBC background model for recovery is parameterized by a power-law index and an amplitude at the reference frequency $ 25\mathrm{Hz} $, such that
	\begin{align}
		\Omega_{\mathrm{rec}}^\mathrm{CBC}(f;\Omega_0, \alpha)\equiv \Omega_0\left(\frac{f}{25\mathrm{Hz}}\right) ^\alpha.
		\label{eq:CBC_rec}
	\end{align}
	We refer to Eq.~\eqref{eq:CBC_rec} as a CBC model, even though this could be generally called ``power-law spectrum model'', because the degrees of freedom in this model can be used to recover the parameters of the CBC injection.
	Also, priors for $\Omega_0$ and $\alpha$ are taken as a uniform distribution in the range of \SIrange{e-10}{e-17}{} and \SIrange{-5}{5}{} respectively, while the priors for $m_b$ and $\chi_\mathrm{ul}$ are identical to those described in Section~\ref{sec:injection}. The parameters considered when evaluating the evidence for each background model are listed in Table~\ref{tab:parameters}.
	
	\begin{figure}
		\includegraphics[width=\linewidth]{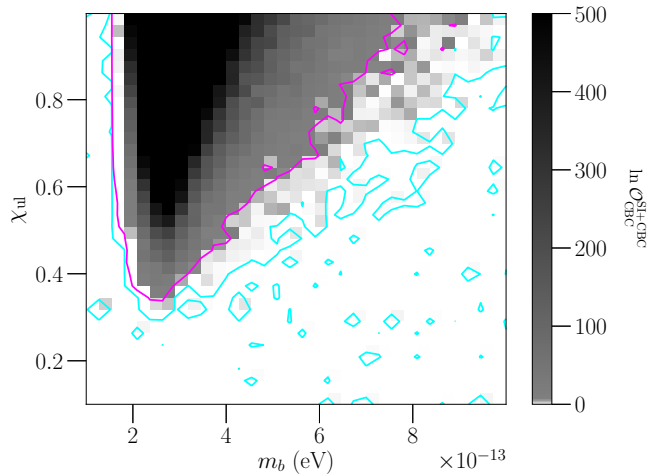}
		\caption{\label{fig:odds_map} Gray-scale map of a log Bayes factor between a superradiant instability $ + $ CBC model and a CBC-only model. The magenta contour represents $ \ln (\mathrm{BF})=8$, while the cyan is $ \ln (\mathrm{BF})=0$.}
	\end{figure}
	We compute a log Bayes factor between these two hypotheses. The computation is repeated, changing injected $ (m_b, \chi_\mathrm{ul}) $ values until we explore a grid over the entire prior space. Thus, we can probe the parameter space on which the superradiant instability signal can be discerned from the expected CBC background with statistical significance. Fig.~\ref{fig:odds_map} is a gray-scale map of a log Bayes factor, effectively equivalent to Eq.~\eqref{eq:ms_BF}, with two contours of $ \ln (\mathrm{BF})=8$ (magenta) and $ 0 $ (cyan). Since the injected CBC background is loud enough to be detected in this simulation, Fig.~\ref{fig:odds_map} implies that inside the magenta contour one can expect the superradiant instability background to be distinguished from the CBC's.
	\begin{table}[h]
		\begin{center}
			\begin{tabular}{|c||c|c|}
				\hline 
				\rule[-1ex]{0pt}{2.5ex} Models & CBC & SI$+$CBC \\ 
				\hline 
				\rule[-1ex]{0pt}{2.5ex} Parameters & $\Omega_0, \alpha$ & $m_b, \chi_\mathrm{ul}, \Omega_0, \alpha$ \\ 
				\hline 
			\end{tabular} 
			\caption{\label{tab:parameters}Parameters in each recovered background model.}
		\end{center}
	\end{table}
	
\section{\label{sec:O1_zerolag} Application to the first observing run of Advanced LIGO}
	\begin{figure}
		\centering
		\includegraphics[width=\linewidth]{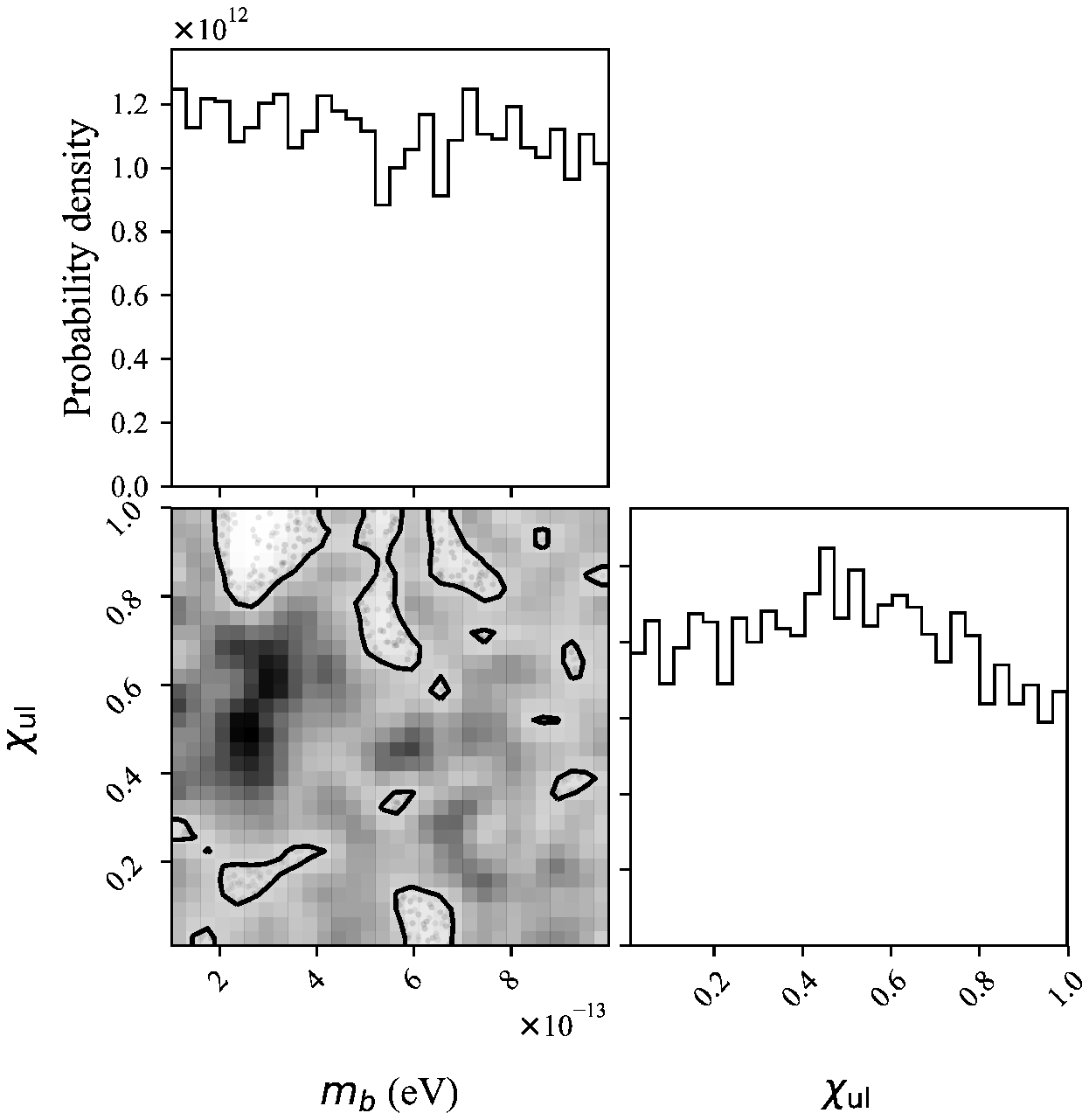}
		\caption{\label{fig:O1_post_ul} Posterior results given by the data from the first Advanced LIGO observing run, recovered with the $ \chi_\mathrm{ul}$ parameterization. The contour on the two-dimensional posterior represents the 95\% confidence level.}
	\end{figure}
	\begin{figure}
		\centering
		\includegraphics[width=\linewidth]{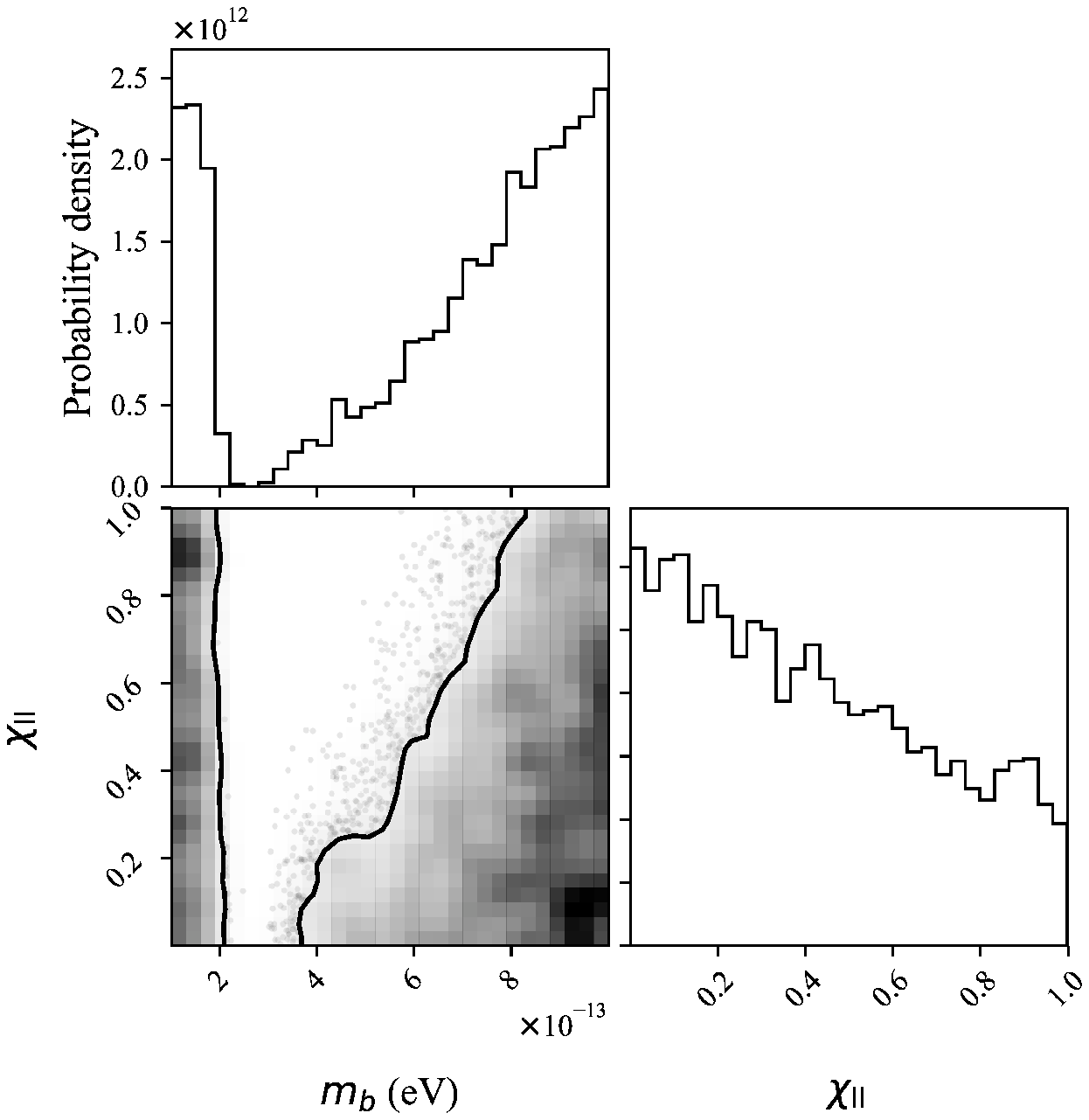}
		\caption{\label{fig:O1_post_ll} Posterior results given by the data from the first Advanced LIGO observing run, recovered with the $ \chi_\mathrm{ll}$ parameterization. The contour on the two-dimensional posterior represents the 95\% confidence level.}
	\end{figure}
	
	We now apply our method to the cross-correlation spectrum measured in Advanced LIGO's first observing run \cite{O1_iso} (the data products given in that paper are publicly available at \cite{O1_release}).
	No statistically significant signal is detected, so we attempt to exclude some of the two-dimensional space of $ m_b $ and $ \chi_{\mathrm{ul},\mathrm{ll}} $ through the search. Figs. \ref{fig:O1_post_ul} and \ref{fig:O1_post_ll} show the posteriors obtained under the $\chi_\mathrm{ul}$ and $\chi_\mathrm{ll}$ parametrizations, respectively.
	Fig.~\ref{fig:O1_post_ul} indicates that, for the $ \chi_{\mathrm{ul}} $ case, the posterior is almost uniformly distributed. Thus, no significant constraint can currently be placed on $m_b$ when $\chi_\mathrm{ul}$ is allowed to vary. This can be also verified by the fact that, according to Fig.~\ref{fig:omega_gw_sum}, almost none of the possible spectra with the $ \chi_\mathrm{ul} $ parameterization can reach the O1 power-law integrated curve. On the other hand, if we fix $\chi_\mathrm{ul}$ and allow the lower bound $\chi_\mathrm{ll}$ to vary, then Fig.~\ref{fig:O1_post_ll} suggests that the mass range $ \SI{2.0e-13}{eV}\leq m_\mathrm{b}\leq\SI{3.8e-13}{eV} $ is excluded. This can be understood by Fig.~\ref{fig:omega_gw_sum} as well, which shows the largest SGWB amplitude when $ m_b\sim\SI{e-12.5}{\electronvolt}\approx\SI{3.2e-13}{\electronvolt}$.
	As a forecast for the future, Appendix~\ref{app_design} shows the constraints that will be possible once Advanced LIGO reaches its design sensitivity.
\section{\label{sec:6}Conclusion}
	This paper presents a first search for signs of supperradiant instability in the SGWB. We describe several performance tests of our search pipeline in Section~\ref{sec:4}. First, a detectable window in scalar mass is estimated from a Bayes factor of injections as shown in Fig.~\ref{fig:BF_plot}. Second, we study the capability to distinguish the model presented here from the fiducial CBC background model. The gray-scale map of log Bayes factors suggests that it is separable from the CBC model with statistical significance in some parameter space (see Fig.~\ref{fig:odds_map}).
	
	Finally, we present results obtained by analyzing data from Advanced LIGO's first observing run.
	No signal is detected with either of our parameterizations of the BH spin distribution, and so we present constraints on possible boson masses and BH spin bounds.
	Using the $ \chi_{\mathrm{ul}} $ parametrization, we cannot place any meaningful constraints on the scalar mass, while the $ \chi_{\mathrm{ll}} $ case rules out the scalar mass range $ \SI{2.0e-13}{eV}\leq m_\mathrm{b}\leq\SI{3.8e-13}{eV} $ with 95\% percent credibility. Let us note that this constraint is still subject to our choice of the BH spin distribution as well as specific astrophysical models we adopt in this work.
	
	Future work will generalize the model to take into account effects of astrophysical uncertainties such as BH formation rate and the initial spin distribution for BBH remnant BHs. Also, the updated local merger rate \cite{O2_pop} should be revisited. As the sensitivity increases with future observing runs of Advanced LIGO and Advanced Virgo, this framework can be used to place stronger constraints on the existence of ultralight bosons.

\begin{acknowledgments}
We thank Richard Brito and Irina Dvorkin for their fruitful discussion and comments. L.T is supported by JSPS KAKENHI Grant Number JP18J21709. LIGO was constructed by the California Institute of Technology and Massachusetts Institute of Technology with funding from the National Science Foundation and operates under cooperative agreement PHY-0757058. A.M and P.M are supported by NSF grant PHY1505870. Parts of this research were conducted by the Australian Research Council Centre of Excellence for Gravitational Wave Discovery (OzGrav), through project number CE170100004. This paper carries LIGO Document Number LIGO-P1800232.
\end{acknowledgments}

\appendix
\section{\label{app_pp} Consistency test of the parameter estimation}
\begin{figure}
	\includegraphics[width=0.9\linewidth]{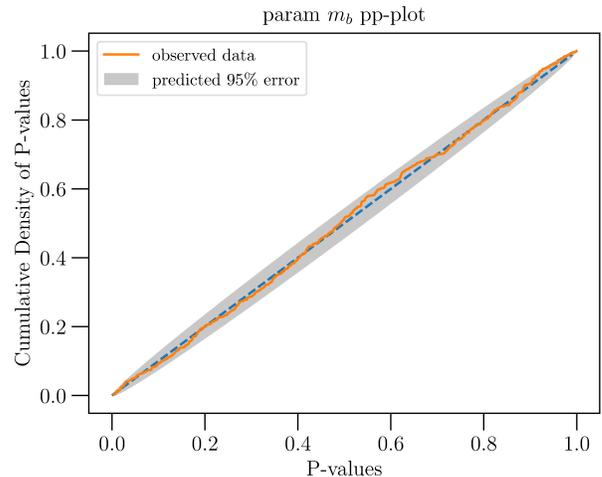}
	\caption{\label{fig:pp-plot} P-p plot obtained by 500 injections into synthesized noise with Advanced LIGO's first observing (O1) sensitivity. Each point represents one injection whose parameters are drawn from prior distributions described in \ref{sec:injection} into different O1 noise realizations. The gray shadow is the $ 95\% $ credible error region predicted by statistical fluctuation from the ideal diagonal. Since the orange curve obtained by data lies within the error region, the pipeline consistently reproduces the parameters.}
\end{figure}

As a statistical consistency check of the parameter estimation performed by our pipeline using \texttt{PyMultiNest}, we construct a probability-probability plot (p-p plot) by performing 500 injections into simulated Gaussian noise drawn from the observed $\sigma(f)$ spectrum from Advanced LIGO's O1 run. For each injection, we first construct the marginalized posterior for $m_b$. We then compute the percentile (or p-value), which is defined by
\begin{align}
	p_i\equiv\int_{m_0}^{m_b^*} p(m_b) \diff m_b,
\end{align} 
where $ i $ denotes the label of a given injection, $ p(m_b) $ the marginalized posterior for $ m_b $, $ m^*_b $ the injected value and $ m_0 $ the lower limit of the $ m_b $ prior space. We then plot the fraction of injections with $p$-values less then a given threshold $ p_i $, as a function of the threshold. For correctly constructed posteriors we expect that for a given threshold $p_i$, the fraction of injections with p-values smaller than the threshold is $p_i$. If this is the case, then the injections will form a straight line with unit slope on the pp-plot. Fig.~\ref{fig:pp-plot} shows the pp-plot we obtain with our injection campaign. The gray shadow is the $ 95\% $ credible error region predicted by statistical fluctuations from the ideal diagonal. Since the orange curve obtained by data lies within the error region, the pipeline consistently reproduces the parameters regardless of the loudness of those injections.

\section{\label{app_design} Demonstration of possible constraints on scalar mass at the design sensitivity}
We run the pipeline with synthesized noise data to demonstrate constraints on scalar mass expected from null result by three years of observation with the Advanced LIGO's design sensitivity. Following the definition Eq.~\eqref{eq:estimator_def} and Eq.~\eqref{eq:sigma_method}, the cross-correlation estimator $ \hat{C}(f) $ and its variance $ \sigma^2(f) $ for individual frequency bins are computed.

\begin{figure}
	\centering
	\includegraphics[width=\linewidth]{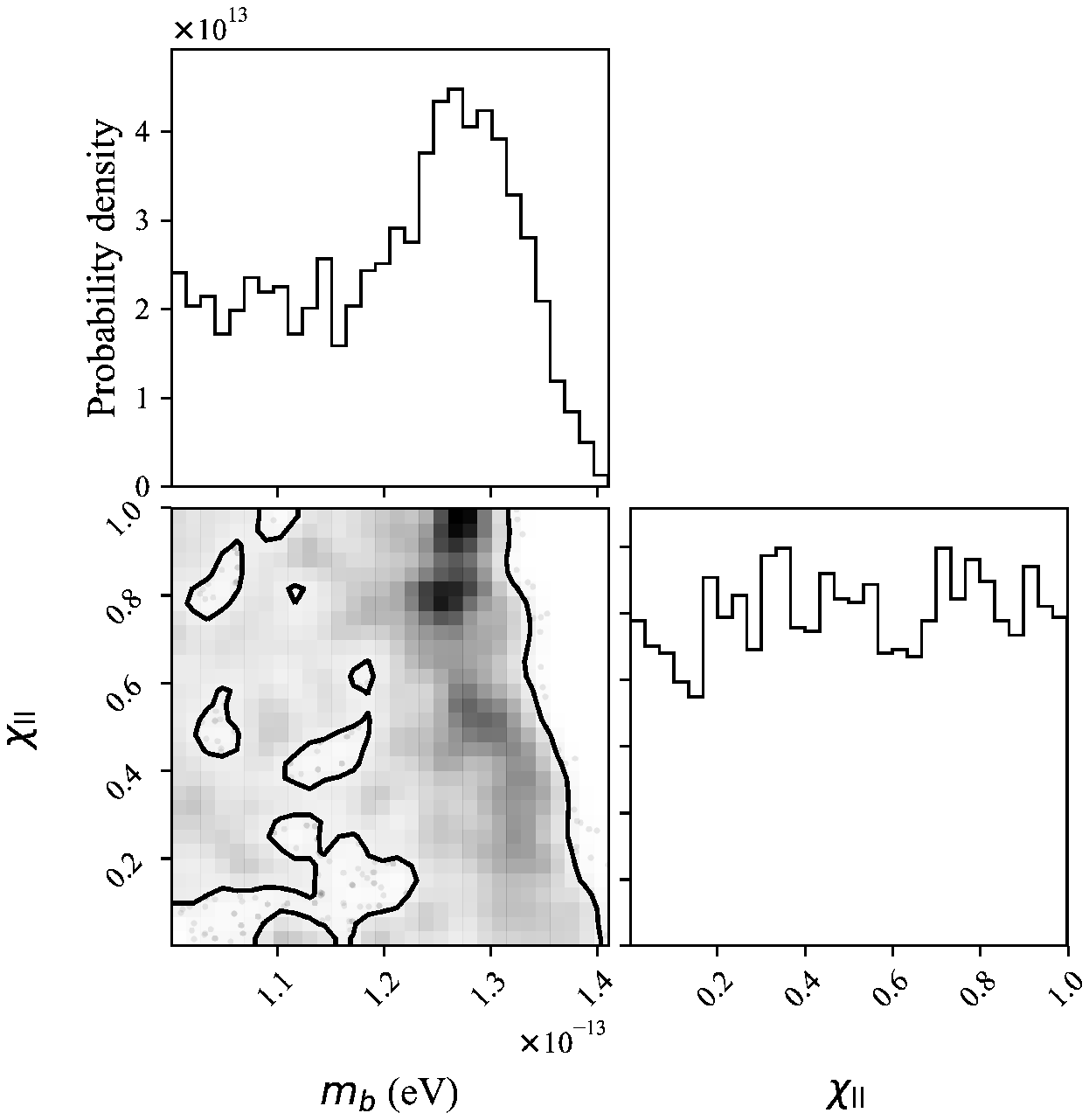}
	\caption{\label{fig:design_post_ll} Posterior results given by a null test for the projected design sensitivity, recovered with the $ \chi_\mathrm{ll}$ parameterization. The contour on the two-dimensional posterior represents the 95\% confidence level.}
\end{figure}
\begin{figure}
	\centering
	\includegraphics[width=\linewidth]{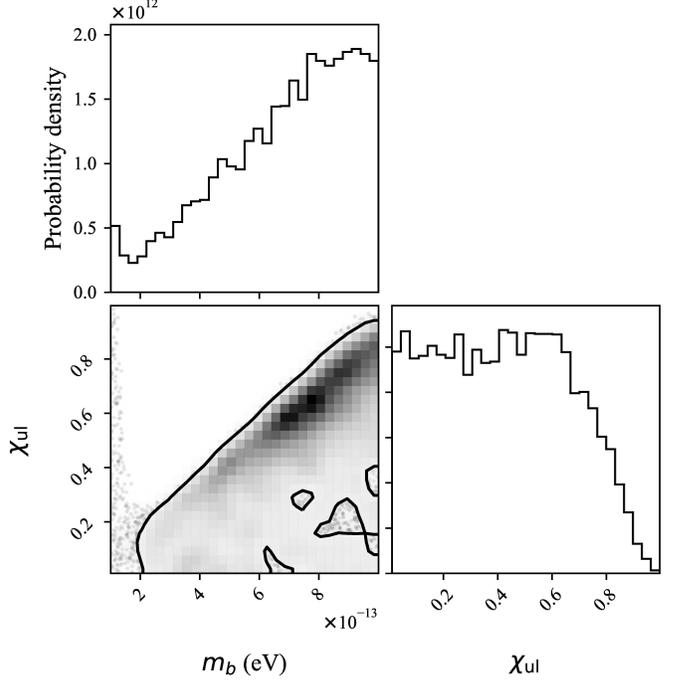}
	\caption{\label{fig:design_post_ul} Posterior results given by a null test for the projected design sensitivity, recovered with the $ \chi_\mathrm{ul}$ parameterization. The contour on the two-dimensional posterior represents the 95\% confidence level.}
\end{figure}

Figs.~\ref{fig:design_post_ll} and \ref{fig:design_post_ul} each shows a two-dimensional posterior with different $ p(\chi) $ parameterizations. When $ \chi_\mathrm{ll} $ is left free, the posterior excludes all scalar masses above $m_b\geq\SI{1.4e-13}{\electronvolt} $, as shown in Fig.~\ref{fig:design_post_ll}. In Fig.~\ref{fig:design_post_ul}, on the other hand, more stringent constraints are placed in lighter scalar mass and the constraints become looser in the heavier boson mass regime. This is due to the strong dependence of the GW background amplitude on $\chi_\mathrm{ul}$. More interestingly, the projected design sensitivity would place quite different constraints depending on the $ p(\chi) $ parameterization. While $ \chi_{\mathrm{ll}} $ parameterization excludes upper part of prior space, $m_b\geq\SI{1.4e-13}{\electronvolt}$, the $ \chi_{\mathrm{ul}} $ parameterization could exclude the lower edge.

\bibliography{bibliography}

\end{document}